\documentclass[aps,prl,twocolumn,showpacs,amsmath,amssymb]{revtex4}

\usepackage{graphicx}
\usepackage{amsmath}
\usepackage{epsf}

\newcommand{\vk}{\mathbf{k}}
\newcommand{\vq}{\mathbf{q}}
\newcommand{\vm}{\mathbf{m}}
\newcommand{\vv}{\mathbf{v}}
\newcommand{\vsigma}{\mbox{\boldmath $\sigma$}}

\newcommand{\vu}{\mathbf{u}}

\newcommand{\tr}{\mbox{tr}\, }

\begin{document}
\title{Mesoscopic bound on anisotropy in itinerant ferromagnets}

\author{Piet W.\ Brouwer and Denis A.\ Gorokhov}
\affiliation{Laboratory of Atomic and Solid State Physics, Cornell
University, Ithaca, New York 14853-2501}

\date{\today}


\begin{abstract}
We calculate the anisotropy energy of a single-domain
ferromagnetic particle in which the
only source of anisotropy is the presence of non-magnetic
impurities. We find that such anisotropy takes the form of 
combined easy-axis and easy-plane anisotropies, with random
orientations of the axes. Typically the anisotropy energy 
is of order $N^{1/2} \hbar/\tau_{\rm so}$, where $N$ is the number
of electrons in the ferromagnetic particle and $\tau_{\rm so}$
is the spin-orbit time.
\end{abstract}

\maketitle

Not all magnetization directions are equal for a single-domain
ferromagnet. Through magnetic dipole interactions and spin-orbit
coupling, the shape of a ferromagnetic particle and the underlying 
crystal lattice lead to preferred directions for the magnetization
\cite{kn:kanamori1963}.
In the absence 
of these anisotropy sources, how isotropic can a ferromagnetic 
particle be? We'll answer this question for an itinerant
ferromagnet with non-magnetic impurities.

For shape or magnetocrystalline anisotropy the energy gain for
a magnetization pointing along a certain `easy axis' or inside an
'easy plane' is proportional to the number of electrons $N$
contributing to the magnetic moment. Clearly, for the question 
asked here, there is no such bulk effect. 
However, for small itinerant ferromagnets, non-magnetic
disorder or irregularities in the boundary of the particle 
break the rotation symmetry on the microscopic scale, thus
creating a `random' anisotropy in the orbital part of electron
wavefunctions. This orbital anisotropy affects the magnetization
through spin-orbit coupling. The resulting `mesoscopic' anisotropy 
does not scale 
proportional to $N$, but it is a fundamental lower bound on 
the anisotropy energy of a ferromagnetic particle. It can be an 
important contribution in
ferromagnetic nanoparticles or in disordered ferromagnetic 
alloys. 

Recently Gu\'eron {\em et al.} \cite{kn:gueron1999} and Deshmukh 
{\em et al.} \cite{kn:deshmukh2001} studied the
excitation spectra of ferromagnetic particles in the nm size
range. The excitation spectrum provides an indirect method to 
measure the magnetic anisotropy energy 
\cite{kn:canali2000,kn:kleff2001,kn:canali2003,kn:cehovin2003}. 
Although shape dominates the magnetic anisotropy in these
nanoparticles, the experiments revealed a significant mesoscopic
contribution to the anisotropy \cite{kn:gueron1999,kn:deshmukh2001}. 
Mesoscopic contributions to the magnetic anisotropy energy have 
also been considered theoretically by Cehovin {\em et al.}
\cite{kn:cehovin2002} and by Usaj and Baranger \cite{kn:usaj2004}.
Motivated by the experiments of Refs.\ 
\onlinecite{kn:gueron1999,kn:deshmukh2001}, these authors studied 
how the addition of a 
single electron changes the magnetic anisotropy energy in a
ferromagnetic particle with a systematic source of uniaxial
anisotropy. The question we ask here is different: it 
is about the mesoscopic lower limit of the magnetic anisotropy when
the magnetic anisotropy energy is dominated by mesoscopic
effects. This may be appropriate for the smallest ferromagnetic 
nanoparticles with low magnetocrystalline
or shape anisotropy. 

We consider an ensemble of ferromagnetic particles, with
equal shape, size and concentration of non-magnetic 
impurities, but with a different configuration of impurities for
each particle. All particles in the ensemble are roughly spherical
and have no notable magnetocrystalline anisotropy. For such an
ensemble, the average free energy $F$ does not depend 
on the direction of the magnetization $\vm$. Hence, the 
magnitude of the `mesoscopic' anisotropy must be characterized
through its sample-to-sample fluctuations, which can
be inferred from the correlation function
\begin{equation}
  C(\theta) = \langle F(\vm_1) F(\vm_2) \rangle
  \label{eq:C}
\end{equation}
where $\theta$ is the angle between the magnetization directions
$\vm_1$ and $\vm_2$. 

In principle, the ferromagnetic
exchange field must be calculated self consistently and at a
fixed number of electrons, taking into account the disorder. We
simplify the calculation by fixing the exchange field a priori
(parallel to $\vm$), and by performing the average at 
a fixed chemical potential $\mu$, rather than a fixed number of
particles. The latter simplification does not
significantly affect the fluctuations of the free
energy \cite{kn:bouchiat1989,kn:cheung1989,kn:altshuler1991}.
The system under consideration is then
described by the single-electron Hamiltonian
\begin{equation}
  {\cal H} = {\cal H}_0  + {\cal V},\ \ 
  {\cal H}_0 = \frac{p^2}{2 m} - E_{\rm Z} \vm
  \cdot \vsigma - \mu,
\end{equation}
where $E_{\rm Z} = \mu_B B_{\rm ex}$ is the Zeeman energy
corresponding the exchange field $B_{\rm ex}$, $\vsigma$ is
the vector of Pauli matrices, and ${\cal V}$
describes the effect of potential scattering and spin-orbit
scattering. In a ferromagnet, $E_{\rm Z}$ is comparable in 
magnitude to the chemical potential $\mu$. 

The free energy $F$ is written as a contribution
$F_0$ in the absence of the potential ${\cal V}$, which is
isotropic, and a shift $\Delta F$. The shift $\Delta F$ is 
calculated as
\begin{eqnarray}
  \Delta F &=& T \sum_{\vk,n} \tr G_0(\omega_n,\vk) {\cal
  V}_{\vk,\vk}
  + \frac{1}{2} T\!\! \sum_{\vk_1,\vk_2,n}\!\! \tr G_0(\omega_n,\vk_1) 
  \nonumber \\ && \mbox{} \times
  {\cal
  V}_{\vk_1,\vk_2} G_0(\omega_n,\vk_2) {\cal V}_{\vk_2,\vk_1}
  + \ldots,
  \label{eq:dF}
\end{eqnarray}
where $\omega_n = \pi T(2n+1)$, $n$ integer, is the fermionic
Matsubara
frequency, $G_0(\omega_n,\vk)$ is $2 \times 2$ matrix Green function
corresponding to ${\cal H}_0$, and ``$\tr$'' indicates a trace over
the spin degrees of freedom. 

The Fourier transform $V_{\vk_1,\vk_2}$ of the potential 
${\cal V}$ is
\begin{eqnarray}
  {\cal V}_{\vk_1,\vk_2} &=&
  v_{\vk_2-\vk_1} - i v^{\rm so}_{\vk_2-\vk_1} \vsigma \cdot
  (\vk_2 \times \vk_1),
\end{eqnarray}
where $v$ and $v^{\rm so}$ are the impurity potential and the
spin-orbit potential, respectively.
For the calculation of the correlator $C(\theta)$ we need to
perform the ensemble
average over the disorder potentials $v$ and $v^{\rm so}$.
We take these to be Gaussian random variables, 
as is appropriate for a small concentration of impurities,
\begin{equation}
  \langle v_{\vq} v_{\vq'} \rangle =
  \frac{w_{\vq}}{V} \delta_{\vq,-\vq'},\ \
  \langle v_{\vq}^{\rm so} v_{\vq}^{\rm so} \rangle =
  \frac{w^{\rm so}_{\vq}}{V} \delta_{\vq,-\vq'}.
  \label{eq:vv}
\end{equation}
The disorder strengths $w_{\vq}$ and $w^{\rm so}_{\vq}$ are
related to the golden-rule scattering times $\tau$ and $\tau_{\rm so}$
for elastic scattering and spin-orbit scattering, respectively.
We assume that disorder is weak and that scattering processes are
predominantly spin-conserving, {\em i.e.,}
$\mu, E_{\rm Z} \gg \hbar/\tau \gg \hbar/\tau_{\rm so}$. No 
assumption is made on the relative magnitude of the
spin-orbit scattering rate $\hbar/\tau_{\rm so}$ and the mean
spacing $\delta$ between energy levels. For ease of presentation
and for technical simplicity, the calculations below are for the 
case of a `half metal', $E_{\rm Z} \gg \mu$, for which all electrons 
in the conduction band have the same spin direction.
For the correlation function of the disorder
potential we take the simple form
\begin{equation}
  w_{\vq} = w \theta(2 k_F - q),\ \
  w_{\vq}^{\rm so} = w^{\rm so} \theta(2 k_F - q),
  \label{eq:simple}
\end{equation}
where $k_F$ is the Fermi wavevector of the half metal and
$\theta(x) = 1$ if $x > 0$ and $0$ otherwise. In this
simple model, the spin-orbit time for electrons at the 
Fermi level is given
by $\tau_{\rm so} = 9 \pi \hbar^3/2 w^{\rm so} k_F^5 m$,
where $\hbar k_F$ is the Fermi momentum.
Although there will be quantitative differences, our 
order-of-magnitude estimates continue to hold for a true itinerant
ferromagnet in which both minority and majority 
electrons are present and if the disorder correlator differs
from the simple model (\ref{eq:simple}).

\begin{figure}
\epsfxsize=0.9\hsize
\epsffile{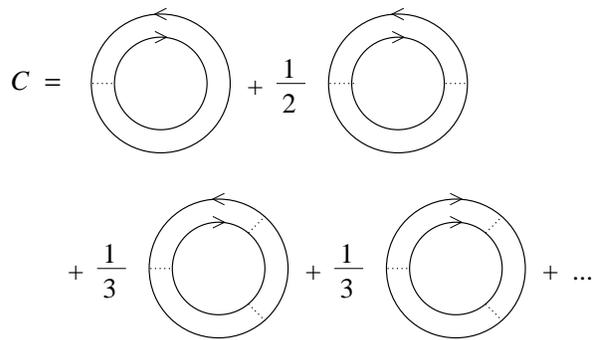}
\caption{\label{fig:1}
Diagrammatic representation of $C(\theta)$. The dotted lines 
represent the impurity average. The solid lines are impurity
averaged single-electron Green functions.}
\end{figure}

Upon performing the disorder average for the correlator
(\ref{eq:C}), one finds that the leading contribution is from 
a series of ring diagrams, as shown in Fig.\ \ref{fig:1}. 
Since there is no spin-orbit coupling for forward scattering,
the first order ring diagram does not
contribute to the mesoscopic anisotropy, whereas the higher
order diagrams do. 

The second order contribution $C_2$ is identical to what
one would find in standard second order perturbation theory
in ${\cal V}$, except for the replacement of $G_0$ by the
impurity averaged Green function. This replacement gives
rise to a smearing of momenta by an amount of order of
the inverse mean free path $1/l$, which is irrelevant for the 
final result. Hence, we find
\begin{eqnarray}
  C_2(\theta) &=&
  \sum_{\vq \neq 0}
  \frac{(w_q^{\rm so})^2}{V^2} \sum_{\vk_1 < k_F}
  \frac{((\vq \times \vk_1)\cdot \vm_1)^2}{
  \varepsilon_{\vk_1} - \varepsilon_{\vk_1+\vq}} 
  \nonumber \\ && \mbox{} \times
  \sum_{\vk_2 < k_F}
  \frac{((\vq \times \vk_2)\cdot \vm_2)^2}{
  \varepsilon_{\vk_2} - \varepsilon_{\vk_2+\vq}},
  \label{eq:dCdtheta2}
\end{eqnarray}
plus terms that do not depend on $\theta$. The summations over 
$\vq$, $\vk_1$, and $\vk_2$ can be done analytically, with the
result (again omitting terms that do not depend on $\theta$)
\begin{equation}
  C_2(\theta) = - N c
  \left( \frac{\hbar \sin \theta}{2 \pi \tau_{\rm so}} \right)^2,
  \label{eq:C2}
\end{equation}
where $c = 3 ( 5005 + 24 \pi^2)/25025 \approx 0.63$ is 
a numerical constant and $N$ is the number of electrons contributing 
to the magnetic moment. Below we show that the 
contribution to $C(\theta)$ of the higher order diagrams
is parametrically smaller than the second-order contribution
$C_{2}(\theta)$. Hence, Eq.\ (\ref{eq:C2}) contains the 
dominant contribution to the correlation function $C(\theta)$.
Higher order contributions are important only for the dependence 
of $C(\theta)$ on an applied magnetic field or temperature
\cite{kn:berkovits}. However, such
dependencies are weak and will not be considered here.

It is important to note 
that the entire Fermi sea contributes to the magnetic anisotropy
fluctuations. This is similar to, {\em e.g.,} (unscreened) 
spontaneous dipole moment of a metal
grain \cite{kn:berkovits}, but different from
most other mesoscopic fluctuation effects, such as persistent currents
and conductance fluctuations \cite{kn:imry2002}, which are determined by
the properties of electronic states around the Fermi level only. As
a consequence, $C(\theta)$ is not universal; it
depends on the details of the impurity 
potential or on the electronic structure of the ferromagnetic
particle. Indeed, in the general case of a ferromagnet with both 
majority and minority electrons, or for a more general disorder
correlator than the simple model (\ref{eq:simple}), the value 
of the numerical constant $c$ will be different, although the
order of magnitude of $C(\theta)$ and the angular dependence 
will be those of Eq.\ (\ref{eq:C2}) above.

We now show that the contribution from the higher order
diagrams to the correlator $C(\theta)$ is small. Hereto we again
consider the case of a half metal, $E_{\rm Z} \gg \mu$. In order
to remove the prefactor $1/n$ from the $n$th order diagram,
$n=3,4,\ldots$, we
calculate $\partial C(\theta)/\partial \theta$ instead of 
$C(\theta)$.
Because $\tau_{\rm so} \gg \tau$, angular averages for the
impurity lines and the single-electron Green functions can 
be performed separately for the higher-order diagrams.
Neglecting the $q$-dependence of the
disorder correlator in Eq.\ (\ref{eq:vv}), we then find
\begin{eqnarray}
  \frac{\partial C}{\partial \theta}
  &=& \frac{\partial C_2}{\partial \theta} +
  \frac{T}{\pi} \sum_{\vq,p > 0} \sum_{\pm}
  \frac{\partial K_{\pm}(\Omega_p,\vq,\theta)}{\partial \theta}
  \nonumber \\ && \mbox{} \times
  \frac{\Omega_p  K_{\pm}(\Omega_p,\vq,\theta)^2}{1 -
  K_{\pm}(\Omega_p,\vq,\theta)},
\end{eqnarray}
where $\Omega_p = 2 \pi k T p$ is the bosonic Matsubara frequency
and
\begin{eqnarray}
  K_{\pm}(\Omega_p,\vq,\theta) &=& 
  \left\langle \frac{\tau_{\rm so}
  - \tau(1 \pm \cos \theta)}{\tau_{\rm so}(1 + \Omega_p \tau - i \hbar \vq \cdot \vk \tau/m)}
  \right\rangle_{\vk}. \nonumber
\end{eqnarray}
Here the brackets $\langle \ldots \rangle_{\vk}$ denote an angular
average over the direction of $\vk$. For small frequencies
$\Omega_p$ and wavevectors $\vq$ one can approximate $K$
as
\begin{equation}
  K_{\pm}(\Omega_p,\vq,\theta) = 1 - \Omega_p \tau 
  - q^2 l^2/3
  - \frac{\tau}{\tau_{\rm so}} (1 \pm \cos \theta).
  \label{eq:Klow}
\end{equation}
When either
$\Omega_p$ or $q$ are large, one can approximate
\begin{equation}
  K_{\pm}(\Omega_p,\vq,\theta) \approx 
  \frac{\tau_{\rm so} - \tau (1 \pm \cos \theta)}{\tau_{\rm so}
  \mbox{max}(\Omega_p\tau,2 q l/\pi)}. \label{eq:Khigh}
\end{equation}
The summation over 
$p$ and $q$ in the region in which both
$\Omega_p \tau \lesssim 1$ and $q l \lesssim 1$ 
can be done using the approximation (\ref{eq:Klow}) for
$K$. The zero mode ($\vq = 0$) gives a weakly singular
contribution to $\partial C/\partial \theta$,
which is small in comparison to $C_2(\theta)$,
whereas the nonzero modes give a non-singular contribution
$\sim V \theta/l^3 \tau_{\rm so}^2$ for small $\theta$.
The remaining part of the
summation over $p$ and $q$ is done with the
help of Eq.\ (\ref{eq:Khigh}) and gives 
a contribution to $\partial C/\partial \theta$ that is a
factor $\sim 1/k_F l \ll 1$ smaller than $C_2$. 

The above calculation fixes the magnitude of the 
mesoscopic anisotropy, but it does not fix the form of the anisotropy.
In fact, both easy plane and easy axis anisotropies with a 
random orientation of the plane or the axis give the same angular
dependence $C(\theta) \propto \sin^2 \theta$ found from 
second order perturbation theory. The full angular dependence
of $F(\vm)$ compatible with $C(\theta) \propto \sin^2 \theta$
is one described by an anisotropy tensor $H$,
\begin{equation}
  F(\vm) =
  \sum_{\alpha,\beta=1}^{3} m_{\alpha} H_{\alpha \beta} m_{\beta}.
  \label{eq:FH}
\end{equation}
Our previous calculation of the free energy correlator has shown
that anisotropy is dominated by second order perturbation theory.
Using second order perturbation theory to calculate the full
tensor $H$ for a half metal, we find
\begin{equation}
  H_{\alpha \beta} = \sum_{\vq \neq 0} \sum_{\vk < k_F}
  \frac{|v^{\rm so}_{\vq}|^2}{\varepsilon_{\vk} -
  \varepsilon_{\vk + \vq}}
  (\vk \times \vq)_{\alpha} (\vk \times \vq)_{\beta}.
  \label{eq:Hsecond}
\end{equation}
A similar but slightly more complicated expression 
holds for the general case.

{}From a statistical point of view, the matrix structure of
$H$ is the same as that of a matrix $H_{\alpha \beta} = 
\sum_{j} u_{j\alpha} u_{j\beta}$,
where the $\vu_j$ are statistically independent vectors. The
elements of such a matrix have Gaussian distributions, the
variances of the diagonal elements being twice as large as the
variance $\sigma^2$ of the off-diagonal elements. Comparing
the correlator $C(\theta) = - 2 \sigma^2 \sin^2 \theta$ one 
obtains from a Gaussian distribution of $H$ with the
microscopic calculation performed above, one finds
$\sigma^2 \sim N \hbar^2 /\tau_{\rm so}^2$.

Changing to principal axes, $H$ can be brought to diagonal form. 
Denoting the
eigenvalues of $H$ by $h_1<h_2<h_3$, with corresponding
eigenvectors $\vv_1$, $\vv_2$, and $\vv_3$, one can rewrite
Eq.\ (\ref{eq:FH}) as
\begin{equation}
  F(\vm) = - h_{a} (\vm \cdot \vv_1)^2 + h_{p} (\vm \cdot \vv_3)^2
  + \mbox{const},
  \label{eq:Fmprincipal}
\end{equation}
where $h_{a} = h_2 - h_1$ and $h_{p} = h_3 - h_2$ are
anisotropy energies for easy axis and easy plane anisotropy,
respectively.
The eigenvalues and eigenvectors
of a real symmetric matrix with random Gaussian elements
have a distribution that is
well known from random matrix theory \cite{kn:mehta1991}: The
orientation of the eigenvectors $\vv_1$, $\vv_2$, and $\vv_3$
is fully random, whereas the distribution of the three
eigenvalues $h_1 < h_2 < h_3$ is
\begin{equation}
  P(h_1,h_2,h_3) \propto
  \prod_{i < j} (h_j - h_i)
  \prod_{i} e^{-h_i^2/4 \sigma^2}.
  \label{eq:Ph}
\end{equation}
This implies that the distribution of $h_a$ and $h_p$ is
\begin{equation}
  P(h_a,h_p) = \frac{8 h_a h_p (h_a + h_p) \sqrt{6 \pi}}{\sigma^5}
   e^{-(h_a^2 + h_a h_p + h_p^2) /6 \sigma^2}.
  \label{eq:Phap}
\end{equation}
In particular, Eq.\ (\ref{eq:Phap}) predicts that the ratio
$r = h_a/h_p$ of easy plane to easy axis anisotropies has the
distribution
\begin{equation}
  P(r) = \frac{27 r (1 + r)}{8 (r^2 + r + 1)^{5/2}} .
\end{equation}
This result is universal and does not depend on the details of
the impurity potential or the electronic structure that affect 
the magnitude of the mesoscopic anisotropy.

It is instructive to compare the magnitude and the angular dependence
of the mesoscopic anisotropy to that of magnetocrystalline
anisotropy. 
For magnetocrystalline anisotropy, the symmetry of the lattice 
often implies a more complicated angular dependence 
than that of Eq.\ (\ref{eq:Fmprincipal}) \cite{kn:kanamori1963}. 
The magnitude of the magnetocrystalline anisotropy scales 
$\propto \hbar N/\tau_{\rm so}$, but 
the numerical prefactor may be small for polycrystalline samples,
disordered ferromagnetic alloys, or ferromagnets with a
close-to-parabolic dispersion relation. A crude comparison of
magnetocrystalline and mesoscopic anisotropy for transition
metal magnetic nanoparticles can be made using the following 
estimates: $\tau_{\rm so}$ can be taken comparable
to that in Cu \cite{kn:piraux1996}, $\tau_{\rm so}
\sim 10^{-12}$s, whereas the magnetocrystalline anisotropy energy 
is $0.04$ meV/atom for Co and significantly smaller for Fe 
and Ni \cite{kn:canali2003}. With these estimates, the mesoscopic 
contribution to
the anisotropy energy should become dominant for Co nanoparticles
with $\lesssim 10^2$ atoms. For Fe and Ni 
mesoscopic anisotropy fluctuations start dominating already
at larger
particle sizes.

Sofar we have dealt with the anisotropy energy $F(\vm)$ at a
fixed number of electrons. This energy determines the size of
the energy barrier for magnetization reversal and the low-energy
particle-hole and magnetic excitations. An important question for 
the experiments of Refs.\
\onlinecite{kn:gueron1999,kn:deshmukh2001} is how much the anisotropy
constants $h_a$ and $h_p$ change upon addition of a single electron to 
the ferromagnetic particle. 
A theoretical estimate of $\delta h_a$ and $\delta h_p$
was given in Refs.\ \cite{kn:cehovin2002,kn:usaj2004}. Both
theories assumed that there was a systematic source of
uniaxial anisotropy, which is appropriate for the non-spherical
nanoparticles used in the experiment; Addition of an electron caused a 
random change $\delta h_a \sim \hbar/\tau_{\rm so} \sim h_a/N$ 
\cite{kn:usaj2004}.
Here we briefly address the same question for 
a ferromagnetic 
nanoparticle with `mesoscopic' anisotropy only. As we'll show
below, both $\delta h_a$ and $\delta h_p$ remain of order $\hbar/\tau_{\rm
  so}$ \footnote{Ref.\ \onlinecite{kn:usaj2004} shows
that $\delta h_a \sim h_a/N$ for very weak spin-orbit scattering,
$\hbar/\tau_{\rm so} \ll \delta$. The argument given below does not
need this assumption.}. However, since the `mesoscopic' anisotropy
is weaker, the relative magnitude of the changes $\delta h_a$ and
$\delta h_p$ is larger: $\delta h \sim h/N^{1/2}$.

The change of anisotropy energy upon addition of a single
electron in an energy level $\varepsilon$ follows from the dependence
of $\varepsilon$ on the magnetization direction $\vm$
\cite{kn:cehovin2002,kn:usaj2004}. 
The $\vm$-dependence of a single level
$\varepsilon$ is well described by random matrix theory. For a
half metal, there are no spin-flip processes, and one finds
that $\varepsilon$ is an eigenvalue of the $M \times M$ hermitian
matrix
\begin{equation}
  {\cal H}(\vm) = {\cal S} + i \sqrt{\frac{\pi \hbar}
  {4 M \tau_{\rm so} \delta}}
  \sum_{\alpha=1}^{3} {\cal A}_{\alpha} m_{\alpha},
  \label{eq:Hm}
\end{equation}
Here $\delta$ is the mean level spacing, ${\cal S}$ is a real symmetric 
matrix, and the ${\cal A}_\alpha$
are real antisymmetric  matrices, $\alpha=1,2,3$. The
elements of all matrices are chosen as random Gaussian variables,
with the same variance for the off-diagonal elements. Diagonal
elements of ${\cal S}$ have a double variance. The matrix size $M$ has
to be taken to infinity. 

In order to find how the addition of a single electron changes
the anisotropy energy $F(\vm)$ for $\vm $ close to the direction
$\vm_0$ at which $F(\vm)$ is minimal, we expand $\varepsilon(\vm)$ 
around $\vm_0$. Statistical estimates of the first and second
derivatives of $\varepsilon$ to $\vm$ are relatively straightforward
because $H(\vm_0)$
and $H(\vm) - H(\vm_0)$ may be considered statistically independent 
for small angular deviations.
The first and second derivatives of $\varepsilon$ to $\vm$ are 
zero on average. 
The mean square average of the first derivative is 
$\langle |\partial \varepsilon/\partial \vm|^2 \rangle = 
\hbar \delta/2 \pi \tau_{\rm so}$. Typically,
the second derivative is of order $\hbar/\tau_{\rm so}$.
Using
Eq.\ (\ref{eq:Fmprincipal}) for $F(\vm)$, with $h_{a}, h_{p} \sim
\hbar
N^{1/2}/\tau_{\rm so}$, one finds that addition of an electron
causes a shift of the preferred magnetization direction
$\vm_0$ by an angle of order $(\tau_{\rm
    so} \delta/N \hbar)^{1/2}$. The shifts $\delta h_a$ and $\delta h_p$
of the anisotropy constants for easy-axis and easy-plane 
anisotropy are $\sim \hbar/\tau_{\rm so}$, in agreement with Refs.\
\onlinecite{kn:cehovin2002,kn:usaj2004}. Note that the expansion of 
$\varepsilon$ around $\vm_0$ is valid up to an
angle $\theta_{\rm max} \sim \min[(\tau_{\rm so} \delta/\hbar)^{1/2},1]$. 
Both the angular shift of $\vm_0$ and the angular distance of
different quantized levels are much smaller than $\theta_{\rm max}$, 
so that the expansion 
used to obtain these estimates is justified. 

In conclusion, we have studied magnetic anisotropy in small
ferromagnetic particles without shape or magnetocrystalline
anisotropy. The presence of non-magnetic impurities and spin-orbit
scattering provides a `mesoscopic' lower bound to the particle's
anisotropy energy. This `mesoscopic' magnetic anisotropy is
caused by spin-orbit coupling, which connects the orbital
anisotropy from non-magnetic impurity scattering to the 
spin degree of freedom. The strength of the `mesoscopic'
anisotropy is of order $N^{1/2} \hbar/\tau_{\rm so}$, where
$N$ is the number of electrons contributing to the magnetic
moment and $\tau_{\rm so}$ is the spin-orbit time, the 
proportionality constant depending on the details of the electronic
dispersion relation and the impurity potential. The angular
dependence of the `mesoscopic' anisotropy is a random combination
of easy-plane and easy-axis anisotropies.

We would like to thank Kathryn Moler, Dan Ralph, and Bertrand
Reulet for discussions. 
This work was supported by the NSF
under grant no.\ DMR 0334499 and by the Packard Foundation.

\end{document}